\documentclass{article}
\usepackage{arxiv}                  

\usepackage{amsmath}                
\usepackage{amssymb}                
\usepackage{amsfonts}               
\usepackage{microtype}              
\usepackage{latexsym}               
\usepackage{textcomp}               

\usepackage{titlesec}               
\usepackage{ulem}                   
\usepackage{enumitem}               
\usepackage{verbatim}               
\usepackage{mdframed}               
\usepackage{booktabs}

\usepackage{array}                  
\usepackage{makecell}               
\usepackage{booktabs}               

\usepackage{graphicx}               
\usepackage{float}                  
\usepackage{caption}                
\usepackage{subcaption}             

\usepackage{gensymb}                
\usepackage{siunitx}                
\DeclareSIUnit\angstrom{\text{Å}}   

\usepackage{hyperref}
\usepackage{xurl}  

\hypersetup{
    colorlinks=true,   
    linkcolor=blue,    
    filecolor=magenta, 
    urlcolor=blue,     
    citecolor=blue,    
    pdfauthor={Your Name},
    pdftitle={Your Paper Title},
}
\usepackage{cleveref}               

\usepackage[numbers,sort&compress]{natbib}
\usepackage{algorithm}              
\usepackage{algorithmic}            
\usepackage{xcolor}                 

\title{The Three Axes of Success: A Three-Dimensional Framework for Career Decision-Making}

\author{
  Meng-Chi Chen \\
  Massachusetts Institute of Technology \\
  \texttt{edchen93@mit.edu}
}

\date{\today}

\begin{document}
\maketitle

\begin{abstract}
Career decision-making is a socio-technical problem: individuals exercise bounded agency while navigating labor market institutions, organizational incentive structures, and information asymmetries that shape feasible trajectories. Existing frameworks optimize along single dimensions—financial returns, work-life balance, or mission alignment—without explicit models for inter-dimensional tradeoffs or temporal dynamics. We propose \textit{The Three Axes of Success}, a normative decision framework decomposing career trajectories into \textit{Wealth} (career capital accumulation and economic optionality), \textit{Autonomy} (control over task selection, temporal allocation, and strategic direction), and \textit{Meaning} (counterfactual social impact scaled by problem importance and personal replaceability). We formalize coupling dynamics between axes: the adjacent possible mechanism by which skill frontiers enable mission discovery, creating nonlinear Wealth$\rightarrow$Meaning transitions; autonomy prerequisites where insufficient career capital triggers control traps; and dual-career household constraints that yield Pareto-suboptimal Nash equilibria under independent optimization. We operationalize each axis through measurable proxies, analyze prototypical career archetypes—industrial R\&D, academia, entrepreneurship—as points in $(W, A, M)$-space, and derive sequential versus simultaneous optimization strategies under uncertainty. The framework converts implicit career anxiety into explicit multi-objective optimization problems with satisficing thresholds, structuring the human-system interaction between individual deliberation and institutional constraints. This provides the first unified decision-theoretic treatment of career success, integrating insights from human capital theory, self-determination theory, and effective altruism into a coherent architecture for rational career design.
\end{abstract}


\section{Introduction}
Individuals face career decisions as multi-period optimization problems under uncertainty, where choices at time $t$ constrain feasible sets at $t+1$ through path-dependent capital accumulation \cite{arthur1989competing, dlouhy2018path}. Classical human capital theory, pioneered by Becker \cite{becker1962investment}, models career investment as maximizing discounted lifetime earnings through skill acquisition and training decisions. Yet empirical evidence shows systematically suboptimal choices when non-pecuniary objectives dominate: individuals routinely sacrifice financial returns for autonomy, purpose, or work-life balance, often without explicit frameworks for evaluating these tradeoffs. For example, Jencks \textit{et al.}~(1988) found that non-monetary job attributes collectively outweigh pay in perceived job quality.\cite{jencks1988good, rosenbaum2016money, express2023salary} Likewise, surveys show that autonomy and career relevance are more strongly related to job satisfaction than earnings. This gap manifests as chronic career anxiety: professionals struggle to articulate whether they should optimize for compensation, autonomy, impact, or some combination thereof, lacking formal tools to evaluate opportunity costs or sequence decisions rationally \cite{chen2021relationship}.

\begin{figure}[H]
\centering
\includegraphics[width=16cm]{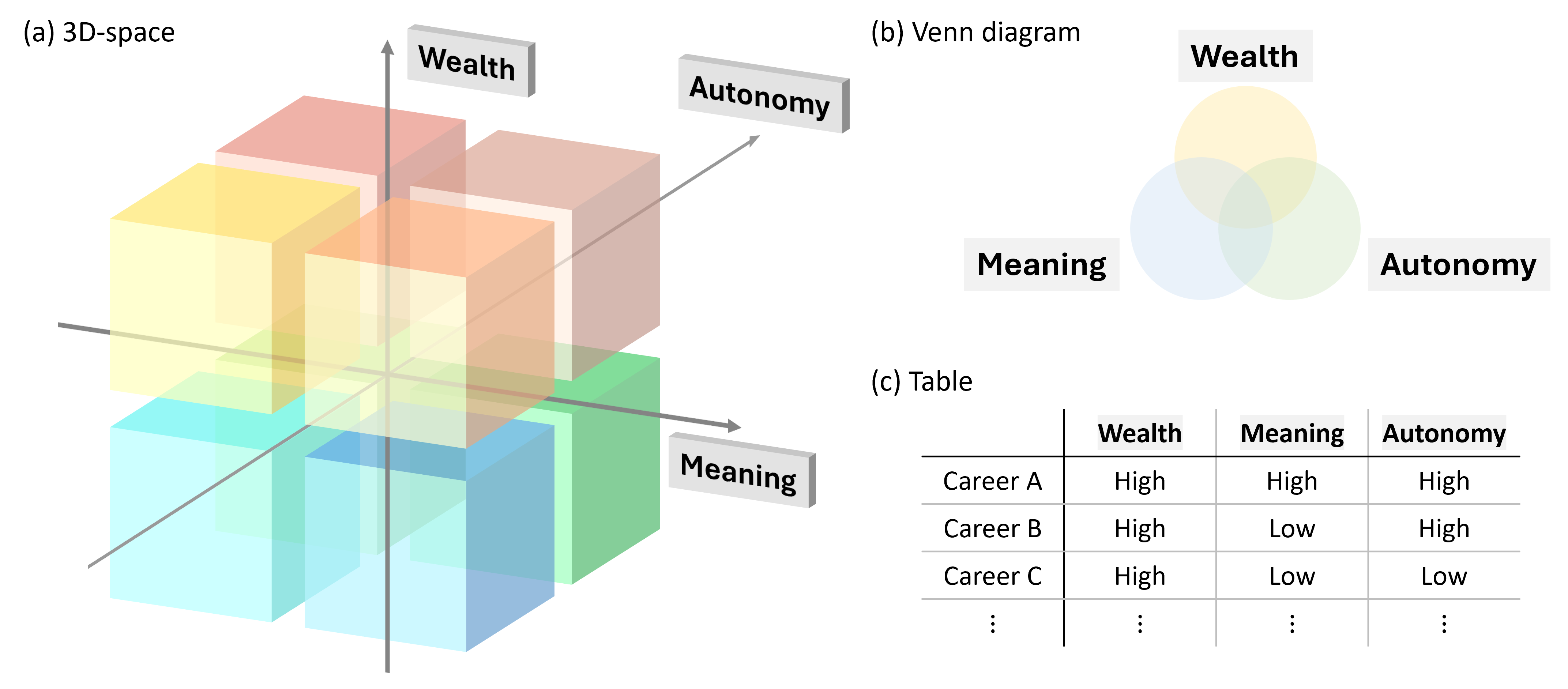}
\captionof{figure}{The Three Axes of Success framework.
(a) Canonical three-dimensional optimization space $(W,A,M)$, which constitutes the formal model used throughout the paper. Career states are represented as points in this space, and feasible trajectories are shaped by path dependence, capital accumulation, and inter-axis coupling.
(b) Two-dimensional Venn-style projection provided solely for conceptual intuition; overlap regions indicate joint satisfaction of value dimensions but do not preserve distances, gradients, or dynamics.
(c) Tabular abstraction illustrating discrete comparisons across careers; this representation is non-spatial and omits coupling and transition costs. Panels (b) and (c) are illustrative projections of (a), not alternative models. Colors are used for visual recognition across panels and serve no analytical purpose.}
\label{3CDM-Fig-01}
\end{figure}

We address this gap by introducing \textbf{The Three Axes of Success}—a three-dimensional constrained optimization framework where axes represent distinct value dimensions: \textit{Wealth} ($W$), \textit{Autonomy} ($A$), and \textit{Meaning} ($M$). We conceptualize career choice as a \textit{socio-technical decision problem}: individuals navigate labor market institutions, organizational incentive structures, and information asymmetries while exercising bounded agency over sequential, path-dependent choices. Decision frameworks that structure this human–system interaction can improve outcomes by making implicit tradeoffs explicit and actionable. We formalize career choice as a multi-objective optimization problem in a three-dimensional value space defined by \textit{Wealth} ($W$), \textit{Autonomy} ($A$), and \textit{Meaning} ($M$), illustrated in Fig.~\ref{3CDM-Fig-01}(a). This three-dimensional representation constitutes the core analytical object of the framework: careers correspond to points or trajectories in $(W,A,M)$-space, and transitions are constrained by path-dependent accumulation of career capital, institutional frictions, and inter-axis coupling dynamics.

For pedagogical purposes, we additionally present two lossy projections of this space. Fig.~\ref{3CDM-Fig-01}(b) provides a Venn-style visualization that highlights joint satisfaction of value dimensions but suppresses metric structure and dynamics. Fig.~\ref{3CDM-Fig-01}(c) offers a tabular abstraction useful for comparative discussion but discards spatial and temporal information entirely. These representations are intended to aid intuition and communication; all formal reasoning and analysis in this paper proceeds from the three-dimensional model in Fig.~\ref{3CDM-Fig-01}(a).

Our framework offers four core contributions:

\begin{enumerate}
\item \textbf{Operational definitions}: We ground each axis in measurable constructs—career capital \cite{newport2016so}, self-determination theory \cite{pink2011drive}, and counterfactual impact \cite{todd202380k}—enabling empirical falsification.

\item \textbf{Coupling dynamics}: We formalize feedback mechanisms where progress on one axis aids or constrains others, including Newport's ``adjacent possible'' (skill frontiers unlock new mission opportunities) and Pink's ``control traps'' (attempting autonomy too early depletes career capital).

\item \textbf{Archetype analysis}: We map canonical career paths (R\&D manager, tenured professor, startup founder) into $(W,A,M)$-space, deriving their Pareto-optimality properties and transition costs.

\item \textbf{Dual-career equilibria}: We extend single-agent models to two-player coordination games, showing how independent optimization by partners creates household-level inefficiencies \cite{rueda2021career}.
\end{enumerate}

The Three Axes of Success framework integrates previously disparate literatures—Becker's human capital theory \cite{becker1962investment}, Pink's self-determination theory \cite{pink2011drive}, Newport's career capital framework \cite{newport2016so}, the 80,000 Hours effective altruism approach \cite{todd202380k}, Simon's bounded rationality \cite{simon1956rational}, and Arthur's path dependence models \cite{arthur1989competing}—into a unified mathematical architecture. We position career archetypes (R\&D manager, tenured professor, startup founder) as points on distinct Pareto frontiers in $(W,A,M)$-space, with predictable tradeoffs and asymmetric transition costs. We analyze labor market scarring effects \cite{ellul2020career} and sequential career strategies \cite{goodlifejourney2025multi} within the framework's coupling dynamics. We synthesize insights from career construction theory \cite{savickas2020career} on narrative identity formation with our quantitative optimization structure.

The remainder of this paper proceeds as follows. Section~\ref{sec:framework} defines the three axes formally, establishing operational definitions and measurement protocols. Section~\ref{sec:coupling} analyzes inter-axis coupling dynamics and temporal dependencies, formalizing the adjacent possible, control traps, and phase-transition logic. Section~\ref{sec:archetypes} characterizes prototypical careers—semiconductor R\&D manager, tenured professor, startup founder—as optimization solutions in $(W,A,M)$-space, deriving their Pareto-optimality properties. Section~\ref{sec:dual} extends the framework to dual-career households, revealing coordination failures and gendered inefficiencies. Section~\ref{sec:strategies} derives decision heuristics under uncertainty, including exploration value, sequential optimization logic, and satisficing thresholds. Section~\ref{sec:related} positions the three-axis framework relative to prior frameworks across human capital theory, motivational psychology, and career planning literature. Section~\ref{sec:conclusion} discusses implications for career counseling, organizational design, and future empirical work.

By converting implicit career anxieties into explicit multi-objective problems, the Three Axes of Success framework enables rational deliberation under value plurality—acknowledging that no single metric captures career success, but that structured reasoning about tradeoffs, coupling dynamics, and sequential constraints improves decision quality across the 80,000-hour working life.

\section{Framework: Three Axes of Career Optimization}
\label{sec:framework}

Career success, we argue, decomposes into three conceptually distinct but empirically coupled objectives: capital accumulation ($W$), temporal and task control ($A$), and counterfactual social impact ($M$). Each axis admits distinct measurement protocols, exhibits characteristic diminishing returns beyond sufficiency thresholds, and interacts with the others through mechanisms formalized in Section~\ref{sec:coupling}. Crucially, the Three Axes of Success framework is \textit{normative} rather than descriptive: the axes represent evaluative objectives for deliberate career design, not empirical claims about how individuals actually make decisions. This positions the framework within the decision-theoretic tradition of multi-objective optimization, where the goal is to structure choices rationally under value plurality, rather than to predict behavior. We acknowledge at the outset that this tripartite decomposition is a modeling choice rather than an empirical discovery; alternative decompositions (e.g., adding health, relationships, or status as separate axes) are defensible and may prove useful in other contexts. Our claim is not that these three axes exhaust the space of career-relevant values, but that they capture the primary tradeoffs documented in the literatures we synthesize, while remaining parsimonious enough for tractable analysis. Formally, at any given time $t$, we represent an individual's career status as a point $x(t) = (W(t), A(t), M(t))$ in $\mathbb{R}^3$. Over the course of a working life, the individual traverses a trajectory through this state space. A particular job or role corresponds to a single point in $(W,A,M)$, and a career move is a transition to a new point with potential gains or losses on each axis. One career outcome (or trajectory) \emph{Pareto-dominates} another if it is at least as high on all three axes and higher on at least one. A career is thus \emph{Pareto optimal} if no other available option yields a higher value in one dimension without reducing another; such careers lie on the efficient frontier of the $(W,A,M)$ space.

\paragraph{Scope boundaries and comparability.} Several clarifications circumscribe the framework's claims. First, the axes are \textit{intrapersonal}: they support comparison of options for a single decision-maker, not interpersonal comparisons of welfare or aggregation across individuals. Second, the numerical representation is \textit{ordinal} for decision purposes—we require only that an individual can rank outcomes along each axis, not that differences are cardinally meaningful. Third, the framework does not claim to model subjective well-being, life satisfaction, or psychological flourishing; these may correlate with $(W,A,M)$ outcomes but are distinct constructs. Finally, transition costs, path dependencies, and feasibility constraints are treated as external to the objective space—they constrain which points are reachable, not the definition of the axes themselves.

Our following subsections provide operational definitions for each axis.

\subsection{Wealth (\texorpdfstring{$W$}{W}): Career Capital and Economic Optionality}

Following Newport \cite{newport2016so}, we define \textit{Wealth} not merely as accumulated income, but as \textit{career capital}—the stock of rare and valuable skills that command market premiums and unlock desirable job characteristics. In this framework, wealth $W$ is a composite stock of career capital that is economically and strategically deployable, encompassing money, skills, actionable knowledge, and leverage-bearing connections. This aligns with human capital theory in economics: Becker (1962) modeled education and training as investments that raise an individual's future productivity and earnings. Consistent with that view, our $W$ axis represents accumulated career investments that increase one's future optionality and economic returns (cf.~\cite{becker1962investment}). Formally, let $W(t)$ denote the market value of an individual’s skill portfolio at time $t$, measurable through:

\begin{itemize}
\item \textbf{Earnings potential}: Total compensation at frontier (90th percentile) roles accessible given current skills \cite{bls2023software, levelsfyi2024frontier}.
\item \textbf{Skill rarity}: Scarcity-adjusted measures of competencies (e.g., H-1B salary premiums, academic citation impact) \cite{peri2012effect, kerr2010supply, garfield1979citation}.
\item \textbf{Optionality}: The number of viable career pivots without capital loss (proxied by industry transferability scores) \cite{lightcast2023labor, burningglassinstitute2023compass}.
\item \textbf{Financial runway}: Liquid assets enabling risk-taking (approximately years of expenses saved) \cite{mmm2012math, vanguard2024has, todd202380k}.
\end{itemize}

Career capital exhibits \textit{diminishing marginal returns} beyond sufficiency thresholds (e.g., \$150k–\$200k household income in high-cost regions stabilizes well-being) \cite{kahneman2010high, todd202380k}, but \textit{increasing returns} in enabling access to high-impact roles that require credentialing or credibility (e.g., tenure-track academia, senior policy positions, founding teams) \cite{todd202380k}. Newport’s central mechanism is that $W$ accumulates through \textit{deliberate practice}—systematically pushing performance beyond current limits with rapid feedback—rather than through tenure alone \cite{newport2016so}.

To capture this accumulation dynamic, we model the rate of career capital growth as a function of both practice quality and existing career capital:

\begin{equation}
\frac{dW}{dt} \propto f(\text{practice quality}, W_{\text{current}})
\end{equation}

This formulation highlights two key features: (i) higher-quality practice (e.g., deliberate practice rather than shallow repetition) accelerates capital accumulation, and (ii) learning is path-dependent, with existing skills amplifying future growth. For example, consider two computer science undergraduates with similar initial $W$ completing internships. One performs repetitive tasks with minimal feedback, while the other works on system-level design problems under senior mentorship with continuous code review. Despite identical pay, the latter experiences a substantially higher increase in $W$ due to superior practice quality.

More generally, this structure implies that individuals can accelerate $W$ accumulation by placing themselves in more demanding environments—those that impose higher standards, stronger feedback, and greater stretch relative to current ability—rather than remaining in low-challenge settings.\cite{mei2021does} The function $f$ may be convex in skill-adjacent domains, where prior expertise accelerates learning, or concave as individuals approach domain-specific skill ceilings \cite{avlonitis2023career}. Importantly, this equation is stylized rather than empirically calibrated; its role is to emphasize path dependence and differential learning rates in career capital accumulation, not to yield quantitative predictions \cite{ericsson1993role, ericsson2004deliberate}.

\paragraph{Non-goals.} The $W$ axis does not claim to measure subjective financial security, consumption utility, or social status. It captures \textit{optionality-weighted market value}—the strategic resource stock that enables future career moves—not the hedonic experience of wealth or its social signaling function.

\subsection{Autonomy (\texorpdfstring{$A$}{A}): Self-Determination and Control}
\label{subsec:autonomy}

Pink's Motivation~3.0 framework\cite{pink2011drive}, rooted in self-determination theory, identifies autonomy—control over task, time, technique, and team—as a primary intrinsic motivator for cognitively demanding work. We operationalize \textit{Autonomy} as the degree of discretionary control over:

\begin{itemize}
\item \textbf{Task selection}: Fraction of work hours allocated to self-chosen projects vs.~assigned deliverables \cite{goodlifejourney2025multi}.
\item \textbf{Temporal sovereignty}: Flexibility in scheduling (remote work, asynchronous collaboration, vacation autonomy) \cite{vanguard2024has}.
\item \textbf{Strategic direction}: Influence over organizational priorities and resource allocation (proxied by managerial span or founder equity) \cite{pink2011drive}.
\item \textbf{Identity alignment}: Freedom to publicly represent professional identity without employer veto (e.g., academic freedom, personal branding) \cite{savickas2020career}.
\end{itemize}

Crucially, Pink \cite{pink2011drive} and Newport \cite{newport2016so} converge on the \textit{control trap}: pursuing autonomy without sufficient career capital ($W$) is unsustainable, due to lack of bargaining leverage or financial runway. Autonomy, in this view, is not a freely choosable preference or personality trait, but a \emph{market-mediated constraint} that becomes feasible only after sufficient capital accumulation. We formalize this dependency as an upper bound on feasible autonomy:

\begin{equation}
A_{\text{feasible}}(t) \leq g(W(t), \text{market structure})
\label{eq-control-trap}
\end{equation}

This inequality states that the maximum level of autonomy an individual can sustain at time $t$ is bounded by a function $g$ of accumulated career capital and prevailing labor-market conditions. For example, at early career stages (e.g., undergraduate internships or junior roles), individuals typically lack the leverage to negotiate remote work, flexible hours, or self-directed projects. Attempting to exceed this bound — such as by prematurely freelancing or demanding extensive flexibility — often results in income instability or forced re-entry into structured employment.\cite{lopes2017determinants}

The function $g$ varies by market structure. In winner-take-all markets (e.g., academia, elite creative fields), autonomy is highly discontinuous and granted only after reaching top performance thresholds. In contrast, auction markets (e.g., software contracting, consulting) exhibit smoother scaling, where autonomy increases gradually with portable, scarce skills \cite{newport2016so}. This formulation abstracts away from important moderators such as industry norms, geographic labor conditions, and individual negotiation skill \cite{lopes2017determinants, nie2023job}, which can be incorporated as secondary parameters.

Finally, autonomy exhibits \textit{threshold effects}: attaining minimal autonomy is often sufficient for retention and motivation, while marginal increases beyond moderate levels show weak correlation with satisfaction \cite{rosenbaum2016money}. This supports treating autonomy as a constrained resource with diminishing returns, rather than a continuously optimizable objective.

\paragraph{Non-goals.} The $A$ axis does not claim to measure psychological feelings of freedom, intrinsic motivation, or job satisfaction. It captures \textit{structural control}—the degree of discretionary authority over work conditions—not the subjective experience of autonomy or its downstream effects on well-being.

\subsection{Meaning (\texorpdfstring{$M$}{M}): Counterfactual Social Impact}

The 80,000 Hours framework \cite{todd202380k} defines career impact not by absolute contribution, but by \textit{counterfactual impact}—the additional good produced relative to the next-best alternative use of one's time and skills. We adopt and operationalize this definition through a multiplicative decomposition:

\begin{equation}
\begin{aligned}
M &= \text{Scale} \times \text{Neglectedness} \times \text{Tractability} \times \text{Personal Fit}\\
\text{Neglectedness} &= \frac{1}{\text{resource currently allocated}}\\
\text{Tractability} &= \frac{\partial (\text{problem solved})}{\partial(\text{effort})}
\end{aligned}
\end{equation}

This equation expresses counterfactual impact as the product of four factors, each of which must be non-negligible for $M$ to be substantial. The multiplicative structure implies that a high score on one factor cannot compensate for a near-zero score on another—a feature that discourages pursuing important but intractable problems or neglected problems where one lacks personal fit. We acknowledge that this formulation embeds specific normative assumptions (e.g., impartial welfarism, aggregation across individuals) that some ethical frameworks would contest.\cite{mcmahan2016philosophical, berkey2021philosophical} The components are defined as follows:

\begin{itemize}
\item \textbf{Scale}: Problem importance (lives affected, welfare magnitude) \cite{todd202380k}.
\item \textbf{Neglectedness}: Inverse of resources allocated (your marginal contribution is larger if fewer people work on the problem) \cite{todd202380k}.
\item \textbf{Tractability}: Feasibility of progress given current knowledge and technology \cite{todd202380k}.
\item \textbf{Personal fit}: Comparative advantage—how much better you are at this problem than alternatives, given your current $W$ \cite{todd202380k}.
\end{itemize}

This framework rejects ``passion-first'' reasoning: effective careers align skills with high-impact problems, not pre-existing interests \cite{newport2016so, todd202380k}. Importantly, $M$ depends on $W$: greater career capital expands the set of tractable, high-impact problems accessible (e.g., AI safety research requires ML expertise; biosecurity policy requires domain credibility).

To illustrate how the multiplicative structure operates in practice, we present three contrasting career opportunities scored on each component. Let $(S, N, T, F) \in [0, 5]^4$ denote ratings for Scale, Neglectedness, Tractability, and Personal fit, respectively.

\begin{itemize}
\item \textbf{Climate advocacy (general awareness campaigns)} | $(S, N, T, F) = (5, 1, 2, 2)$: 
Climate change scores maximally on Scale due to global, long-term consequences. However, the space is heavily resourced with funding, talent, and public attention, yielding low Neglectedness. Tractability for awareness campaigns—as opposed to technical mitigation—is limited, and moderate Personal fit results in a low overall $M$ despite the problem's importance.

\item \textbf{Lead contamination testing (underserved housing)} | $(S, N, T, F) = (2, 4, 4, 4)$: 
Scale is moderate—fewer individuals affected than global challenges—but Neglectedness is high because few actors prioritize this issue. Tractability is strong: established measurement and remediation protocols exist. For individuals with chemistry, public health, or fieldwork expertise, high Personal fit amplifies counterfactual impact, yielding substantial $M$ despite modest absolute scale.

\item \textbf{AI safety interpretability research (internship)} | $(S, N, T, F) = (5, 3, 3, 4)$: 
Scale is maximal given civilizational stakes. Neglectedness is moderate—growing attention, but still undersupplied relative to importance. Tractability is constrained by technical difficulty. Personal fit is decisive: individuals with strong machine learning backgrounds contribute meaningfully, while those lacking such expertise face near-zero marginal impact regardless of motivation.
\end{itemize}

Unlike Wealth $W$ and Autonomy $A$, which yield personal returns, Meaning $M$ is \textit{altruistic} by construction—its value accrues primarily to beneficiaries rather than the agent. Individuals differ in how heavily they weight $M$ relative to self-interested objectives, a heterogeneity we capture through an altruism parameter $\lambda_M \in [0, 1]$. This allows us to express career utility as a weighted objective:

\begin{equation}
U = \lambda_W \cdot W + \lambda_A \cdot A + \lambda_M \cdot M, \quad \text{where } \lambda_W + \lambda_A + \lambda_M = 1
\label{eq:utility}
\end{equation}

Here, $\lambda_W$, $\lambda_A$, and $\lambda_M$ represent the relative importance an individual places on each axis. An agent with $\lambda_M = 0$ optimizes purely for personal returns (wealth and autonomy), while $\lambda_M \approx 1$ reflects strong impartiality characteristic of effective altruist reasoning \cite{zhang2023impacts, todd202380k}. This parameterization explains why individuals with similar career capital pursue divergent paths: a high-$\lambda_M$ software engineer may accept lower compensation for AI, while a low-$\lambda_M$ counterpart maximizes $W$ in fintech. The framework thus accommodates value pluralism without prescribing which weighting is "correct"—it merely makes the tradeoffs explicit.

\paragraph{Non-goals.} The $M$ axis does not claim to measure subjective sense of purpose, existential meaning, or personal fulfillment. It captures \textit{counterfactual social impact}—the marginal contribution to outcomes that matter from an impartial perspective—not the felt experience of meaningful work or its role in identity formation.

\section{Coupling Dynamics: Inter-Axis Dependencies}
\label{sec:coupling}

Although we have defined $W$, $A$, and $M$ as conceptually distinct axes, they exhibit asymmetric causal dependencies in practice. Most critically, Wealth accumulation gates access to both Autonomy and Meaning: insufficient career capital constrains the feasible set of autonomous arrangements and limits the high-impact opportunities one can credibly pursue. Conversely, high-$M$ roles can stabilize Autonomy even at lower Wealth levels by substituting mission-alignment for financial compensation. This section formalizes these coupling dynamics, with particular attention to Newport's ``adjacent possible'' mechanism and the ``control trap'' identified independently by Pink and Newport.

\subsection{The Adjacent Possible: \texorpdfstring{$W \rightarrow M$}{W to M} Coupling}

Newport's \textit{adjacent possible} concept \cite{newport2016so} describes how meaningful missions emerge only after reaching the frontier of one's field—once an individual has accumulated sufficient career capital $W$ (the stock of skills, credentials, reputation, networks, and institutional access) to perceive opportunities invisible to those earlier in their careers. We formalize this idea as a monotonicity condition on the set of accessible high-impact missions, denoted by $\mathcal{S}(W)$:

\begin{equation}
W_2 > W_1 \implies \mathcal{S}(W_2) \supseteq \mathcal{S}(W_1)
\end{equation}

Here, $\mathcal{S}(W)$ denotes the set of missions that are accessible at a given level of career capital, and the subscripts index different points in time. This relation states that if an individual possesses more career capital at a later time ($W_2$) than at an earlier time ($W_1$), then every mission that was previously accessible remains available, and additional opportunities may become feasible. In other words, higher career capital strictly expands—or at minimum preserves—the set of high-impact problems one can meaningfully address.

The intuition is that domain expertise, credentials, and professional networks open doors to opportunities—such as research collaborations, policy positions, or founding roles—that remain invisible or inaccessible at earlier career stages. We note that this monotonicity assumption may fail in cases of extreme specialization (where deep $W$ in one domain forecloses options in others) or rapid technological disruption (where legacy skills depreciate) \cite{de2002economics, blechinger1996technological}.

This creates a \textit{sequential dependency}: attempting to optimize $\mathcal{S}$ prematurely (before acquiring sufficient career capital $W$) tends to result in low-impact roles due to lack of leverage. For example, an undergraduate deeply motivated by climate change may wish to ``work on the most important problems'' immediately and therefore skip skill-building in favor of pursuing high-level policy or advocacy roles. In practice, limited technical expertise, credentials, and professional credibility often restrict such individuals to peripheral or symbolic positions with little influence over real decisions. By contrast, the same individual after several years of domain-specific training, research, or industry experience retains access to those early roles while also becoming eligible for substantially higher-leverage missions \cite{mello2025career}.

Similar dynamics appear in other domains: 
\begin{itemize}
\item Junior policy roles contribute minimally to AI governance compared to senior technical advisors with ML expertise \cite{todd202380k}.
\item Academic missions (e.g., opening new research directions) are inaccessible without tenure-track credentials and citation capital \cite{kerr2010supply}.
\end{itemize}

The 80,000 Hours framework \cite{todd202380k} operationalizes this through “career capital before direct work”: individuals should frontload skill acquisition in high-growth domains (ML, synthetic biology, policy analysis) before attempting high-stakes roles. The transition from $W$-maximization to $M$-maximization is \textit{nonlinear}—there exist critical thresholds (e.g., PhD completion, senior engineering level) where $\partial M / \partial W$ discontinuously increases.

\subsection{Control Traps: \texorpdfstring{$W \leftrightarrow A$}{W and A} Feedback}

Pink \cite{pink2011drive} and Newport \cite{newport2016so} independently identify a two-sided failure mode in the pursuit of autonomy. Both argue that there exists a limited career time window during which accumulated career capital ($W$) can be successfully converted into increased autonomy ($A$). Attempting this conversion too early or too late leads to distinct control traps, described below.

\paragraph{First Control Trap ($A$ without $W$):}
Seeking autonomy prematurely (e.g., freelancing, lifestyle businesses, early entrepreneurship) without sufficient career capital fails due to lack of market leverage, as discussed in Section~\ref{subsec:autonomy}. Clients will not pay premium rates for undifferentiated skills, and individuals lack sufficient financial runway to absorb income volatility. In contrast to Inequality~\eqref{eq-control-trap}, this failure mode can be expressed as the following conditional implication:

\begin{equation}
A_{\text{attempted}} > g(W_{\text{current}}) \implies \text{financial instability}
\end{equation}

This formalizes the intuition that attempting to secure a level of autonomy ($A_{\text{attempted}}$) exceeding what one’s current career capital can support results in unsustainable outcomes. The threshold function $g(\cdot)$ represents the maximum autonomy the labor market is willing to “fund” given demonstrated value. Empirically, this manifests as failed freelance ventures, underpaid consulting, or premature entrepreneurship that consumes savings without building durable income streams \cite{casalhay2025gig, peetz2021living}.

\paragraph{Second Control Trap ($W$ enables $A$, but employers resist):}
At later career stages, accumulating substantial career capital makes individuals increasingly valuable to their employers, who then rationally resist autonomy-granting arrangements or departures. For example, consider a senior software engineer at a large technology firm who becomes the sole expert maintaining a mission-critical internal system. Although their skills are highly marketable and external firms would offer both higher compensation and greater flexibility, internal management denies requests for remote work or reduced scope of obligation precisely because the individual is too important to lose. This second failure mode can be formalized as:

\begin{equation}
W_{\text{current}} \geq W^* \;\wedge\;\ A_{\text{requested}} > A_{\text{granted}}
\;\Longrightarrow\;
\text{organizational resistance}
\end{equation}

Here, $W^*$ denotes the second control-trap threshold beyond which the individual’s marginal productivity becomes difficult to replace internally. Even though $W$ is now sufficient to support autonomy in the external market, internal organizational incentives bind $A_{\text{granted}}$ below $A_{\text{requested}}$. The constraint is no longer market viability, but surplus extraction by the employer.

Newport’s ``law of financial viability'' \cite{newport2016so} provides a diagnostic test: autonomy is viable only if independent buyers are willing to pay for the output. If such buyers exist, $W$ is sufficient; if autonomy is still denied internally, the constraint is organizational rather than economic.

Taken together, these two traps imply that autonomy must be \textit{earned} through $W$ accumulation, but that the conversion from $W$ to $A$ is inherently adversarial. Employers have incentives to appropriate the surplus generated by high-$W$ employees by restricting autonomy. As a result, the feasible $W \rightarrow A$ transition occurs only within a narrow temporal window.

Optimal strategies therefore emphasize:
\begin{itemize}
\item Accumulating $W$ in transferable, externally valued skills \cite{nagele2016education} (e.g., programming, writing, specialized technical expertise) rather than firm-specific capital whose value collapses outside the current organization \cite{dlouhy2018path}.
\item Extracting autonomy incrementally over time—through periodic gains such as partial remote work, side projects, or expanded scope of control—within a single firm rather than attempting a single abrupt transition to full independence \cite{pink2011drive}.
\end{itemize}

\subsection{Meaning as Autonomy-Preserver: \texorpdfstring{$M \rightarrow A$}{M to A} Stability}

High-$M$ roles (nonprofit leadership, academic research on neglected problems, policy work) often trade lower $W$ for sustained $A$: mission-driven organizations grant autonomy to attract talent despite below-market compensation \cite{todd202380k, jencks1988good}. This creates a stabilizing feedback that we express schematically:

\begin{equation}
M \uparrow \implies A \text{ sustainable despite } W \downarrow
\end{equation}

This notation indicates that increases in perceived meaning can substitute for financial compensation in maintaining autonomy. The mechanism is that mission-driven sectors face competitive pressure to offer non-pecuniary benefits (research freedom, flexible hours, value alignment) to compensate for salary gaps relative to commercial alternatives \cite{borzaga2006worker}.

For example, tenured professors accept lower income than industry equivalents in exchange for research autonomy and mission alignment \cite{bls2023software}. However, this equilibrium is fragile: excessive $W$ sacrifice undermines financial viability (the first control trap), particularly for dual-career households.

\subsection{Temporal Dynamics: Phase Transitions in Optimization}

Optimal career trajectories exhibit \textit{phase transitions}—qualitative shifts in which axis is prioritized as constraints relax and opportunities evolve \cite{super1973career, hartung2013life}:
\begin{enumerate}
\item \textbf{Phase I (Years 0–10):} Maximize $dW/dt$ through deliberate practice and credentialing. Defer $A$ and $M$ optimization. \textit{Example:} PhD training, junior engineering, consulting roles \cite{newport2016so, todd202380k}.
\item \textbf{Phase II (Years 10–25):} Leverage $W$ to negotiate $A$ (transition to autonomy-rich roles) or pivot to $M$ (enter high-impact domains now accessible). \textit{Example:} senior individual-contributor positions, founding a startup, tenure-track faculty \cite{pink2011drive, todd202380k}.
\item \textbf{Phase III (Years 25+):} Maintain satisficing $W$ and $A$ while maximizing $M$ (legacy-building). \textit{Example:} distinguished research appointments, policy leadership, mentorship roles \cite{newport2016so}.
\end{enumerate}

We emphasize that these phase boundaries are illustrative rather than universal; they depend on individual circumstances including family obligations, health, prior wealth, and industry-specific career ladders. The phase durations given (0–10, 10–25, 25+) are rough central tendencies from Western professional contexts and may differ substantially across cultures, industries, and cohorts.

The phase boundaries are individual-specific and depend on diminishing returns: once $W$ exceeds sufficiency (e.g., \$150k–\$200k income), once $A$ reaches moderate levels, and once $M$ satisfies personal values, further optimization offers little marginal utility \cite{jencks1988good}.

\section{Career Archetypes in \texorpdfstring{$(W, A, M)$}{(W, A, M)}-Space}
\label{sec:archetypes}

To ground the abstract $(W, A, M)$ framework in recognizable career trajectories, we analyze three canonical archetypes that occupy distinct regions of the optimization space. These archetypes—ind management, tenured academia, and venture entrepreneurship—represent different local Pareto frontiers with predictable tradeoff structures and asymmetric transition costs. We emphasize that these are stylized ideal types, not exhaustive categories; many careers blend elements across archetypes or follow hybrid trajectories. The value of archetype analysis lies in making explicit the implicit weightings and constraints that characterize broad classes of career choices.

\subsection{Archetype Definitions}

We analyze three prototypical careers as points in $(W, A, M)$-space:

\begin{description}[leftmargin=0cm,labelsep=0cm,align=left]
\item[\textit{Industrial R\&D Management ($W^{\uparrow}, A^{\leftrightarrow}, M^{\downarrow}$):}] A career maximizing financial capital and moderate autonomy at the expense of direct social impact. \textit{Profile:} engineering or corporate management roles in large firms (e.g., semiconductor R\&D director). Such roles offer high compensation and leadership control (project choice, team direction), but the product impact is channeled through profit-oriented objectives rather than altruistic missions.
\item[\textit{Tenured Academia ($W^{\downarrow}, A^{\uparrow}, M^{\uparrow}$):}] A career maximizing autonomy and societal impact at the expense of wealth. \textit{Profile:} professor or research scientist in academia or non-profit institutions. These roles offer substantial control over research agenda and contribute to fundamental knowledge or public-good problems, but typically at below-industry pay scales.
\item[\textit{Venture Entrepreneurship ($W^{\uparrow\uparrow}, A^{\uparrow}, M^{\uparrow}$):}] An attempted maximization of all three axes via high-risk, high-reward paths. \textit{Profile:} startup founder or early-stage entrepreneur aiming for outsized financial success ($W$) and autonomy ($A$) while pursuing an impactful innovation ($M$). Such paths can produce Pareto-superior outcomes (e.g., mission-driven companies that achieve unicorn valuations), but also carry high variance and potential collapse (simultaneous optimization is fragile). Importantly, the expected value of entrepreneurial paths is heavily right-skewed: median outcomes are substantially worse than mean outcomes, and survivorship bias in entrepreneurship narratives may lead observers to overestimate success probabilities. \cite{cantamessa2018startups, bipventures2025powerlaw}.
\end{description}

These archetypes inhabit different regions of the $(W, A, M)$ decision space and lie on distinct local Pareto frontiers. Each represents a solution to a multi-objective optimization under implicit weightings: industrial management heavily weights $W$, academia heavily weights $A$ and $M$ (with near-zero $\lambda_W$ beyond a sufficiency threshold), and entrepreneurship aspires to large $W$, $A$, \textit{and} $M$ (effectively assuming an optimistic non-convex frontier).

\subsection{Tradeoff Analysis and Transitions}
The archetypes illuminate inherent tradeoffs. Industrial R\&D careers achieve financial security and mid-level autonomy but often report lower personal meaning, leading to mid-career pivots toward more mission-oriented roles\cite{chen2021relationship}. Academic paths prioritize impact and freedom but face financial constraints; many academics exit to industry post-PhD or mid-career (``post-tenure burnout'') when $W$ pressures mount \cite{savickas2020career, zhang2022teacher}. Entrepreneurial paths promise all-axis success but frequently revert to corporate roles or financial insecurity if ventures fail, demonstrating the risk of simultaneous optimization. \cite{cantamessa2018startups, berglann2011entrepreneurship}

Transitioning between archetypes incurs costs that are often asymmetric. Moving from an industrial role to academia ($W \downarrow, A \uparrow, M \uparrow$) typically requires a substantial pay cut and credentialing (e.g., pursuing a PhD) with uncertain success; moreover, the transition window may be time-limited as research productivity norms favor early-career entry. \cite{rueda2021career, dlouhy2018path}. Shifting from academia to industry ($W \uparrow, A \downarrow, M \downarrow$) may restore earnings but at the expense of purpose and autonomy, leading to potential dissatisfaction \cite{rosenbaum2016money}. Entrepreneurial forays from either academia or industry entail opportunity costs and risk of failure, but can also yield transformative $W,A,M$ outcomes in rare cases.

In practice, many professionals pursue \textit{sequential} archetypes: e.g., an individual maximizes $W$ in early industrial roles, then transitions to a high-$M$ academic or policy position once financially secure. Our framework’s $(W,A,M)$ mapping enables explicit evaluation of such sequences, estimating gains and losses along each axis when moving between career regimes.

\section{Dual-Career Dynamics and Household Optimization}
\label{sec:dual}

The preceding analysis treats career optimization as a single-agent problem. However, for individuals in committed partnerships, career decisions become interdependent: geographic constraints, childcare responsibilities, risk preferences, and income floors create externalities between partners' $(W, A, M)$ trajectories. When partners optimize independently, the household may settle into Pareto-suboptimal equilibria that neither would choose under full coordination. This section extends the framework to two-player settings, identifying common coordination failures and cooperative strategies that approach household-level efficiency. \cite{tam2023migration, yu2025dollars} 

\subsection{Coordination Problems in Dual Careers}
Dual-career couples face additional constraints: geographic co-location, childcare responsibilities, and correlated income risk. In our three-axis model, each partner $i \in {1,2}$ has $(W_i, A_i, M_i)$ trajectories that become interdependent once partnered. Without coordination, the household’s combined objective can fall into a Pareto-suboptimal equilibrium.

A common outcome is an asymmetric equilibrium where one partner sacrifices $W$ and $A$ growth (e.g., taking part-time work or stagnating in a less demanding job) to support the other's career advancement. Empirically, this pattern disproportionately affects women in heteronormative households, contributing to persistent gender gaps in senior positions. \cite{socap2020dualcareer, averkamp2024decomposing} We note that this asymmetry may be individually rational (given prevailing wage gaps and social expectations) even when household-suboptimal, creating a collective action problem that individual couples cannot easily escape. \cite{fernandez2025gender, becker1965theory} The high-$W$ partner maximizes personal career gains assuming support on the home front, effectively consuming household $A$ resources (flexibility, time) supplied by the other partner, whose own $W,A$ potential is curtailed.

Another inefficiency arises from risk correlation: if both partners pursue high-variance, entrepreneurial paths simultaneously, the household's financial stability (aggregate $W$) may fall below acceptable thresholds. Without coordination, partners may either both avoid risk (foregoing potential $W$ gains and $M$ impact) or both embrace it (amplifying downside risk). This phenomenon is analogous to correlated asset returns in portfolio theory, where diversification benefits require negatively correlated or uncorrelated investments. \cite{jermann2002international, blundell2016consumption}

\subsection{Cooperative Solutions}

Optimizing dual careers requires treating the household as the unit of analysis, with a joint objective (e.g., maximizing a weighted sum of $W_1+W_2$, $A_1+A_2$, $M_1+M_2$ or ensuring each exceeds certain thresholds). Cooperative strategies to approach Pareto-optimal frontiers include:

\begin{itemize}
\item \textbf{Sequential focus:} Alternate periods of $W$-maximization between partners. For example, one partner prioritizes career (maximizing $W$ and $A$) while the other temporarily emphasizes household responsibilities ($A \downarrow$ professionally), then switch. Over a lifecycle, both partners achieve high $W$ and $A$ albeit sequentially rather than concurrently.
\item \textbf{Risk hedging:} Stagger risk-taking so that one partner maintains a stable high-$W$ role while the other undertakes a high-variance, high-$M$ or high-$A$ endeavor (e.g., startup or sabbatical). This preserves a baseline $W$ and benefits from upside if the risky venture succeeds.
\item \textbf{Geographic bundling:} Jointly select locations or opportunities that maximize combined $W$ given co-location (e.g., two-body academic hiring, or choosing a tech hub city with opportunities for both). \cite{rueda2021career, yates2022getting} This may involve one partner taking a slightly lower $W$ or $A$ role than individually optimal to enable the other's placement, but the net household outcome dominates separate optima in different cities.
\end{itemize}

These cooperative solutions echo concepts from household economics and game theory, highlighting that individually rational career moves can be collectively suboptimal. By mapping both partners’ careers in $(W,A,M)$ space, the framework facilitates transparent discussions and planning: partners can visualize how a move by one affects the other's feasible set and jointly navigate toward a desired household-level vector (e.g., achieving a minimum combined $W$ while maximizing the sum of $A$ and $M$).

\section{Decision Strategies Under Uncertainty}
\label{sec:strategies}

Career decisions unfold under substantial uncertainty: individuals have imperfect knowledge of their own preferences, future labor market conditions, and the outcomes of specific choices. Classical optimization—selecting the path that maximizes expected utility over $(W, A, M)$—is often intractable and may be mis-specified given bounded rationality. Drawing on Simon's satisficing framework \cite{simon1956rational} and real options theory, this section develops decision heuristics that acknowledge uncertainty while remaining actionable. We argue that exploratory early-career strategies and satisficing thresholds typically dominate single-path optimization under realistic assumptions about information constraints and preference evolution.

\subsection{Option Value of Skill Acquisition}

Early-career decisions should maximize \textit{option value}—keeping future pathways open—given uncertainty in personal fit and evolving opportunities. \cite{jacobs2007real, kemna1993case}. In our framework, building generalizable $W$ (portable skills, broad networks) provides a real option to pivot toward higher $A$ or $M$ later. This aligns with 80,000 Hours' recommended \textit{exploration phase} (first 5–10 career years) \cite{todd202380k} and Newport's \textit{career craftsman} ethos \cite{newport2016so}, wherein one tries multiple domains and hones transferable skills before committing to a specialization.

Formally, consider two skill-building strategies from $t=0$ to $t=T$: one specialized (maximizing $W$ in a narrow domain) and one generalized (investing in diverse skills). The specialized strategy may yield higher $W(T)$ in expectation, but with higher variance across possible $M$ outcomes (limited pivot options if the chosen domain’s impact saturates). The generalized strategy likely yields a slightly lower $W(T)$ but a larger support of $\mathcal{M}(W)$ accessible at $t=T$ (more adjacent possible opportunities). Under uncertainty about personal interests and external impact opportunities, a risk-neutral planner might still choose the generalized path if the option value (expected maximal $M$ attainable by switching later) outweighs the near-term $W$ gain of specialization.

An illustrative case is a new graduate deciding between a specialized PhD program vs.~a rotation program in industry. The PhD might maximize $W$ in academia but if academic $M$ opportunities collapse (e.g., funding cuts or low discovery rate), the person has lower lateral mobility.\cite{Mangematin2000, BenjaminMilouchevaVigezzi2025} The rotation program builds broader $W$ (multiple domains) and keeps both industry and academia open, hedging against uncertainty in both personal fit and external demand.

\subsection{Sequential vs. Simultaneous Optimization}

Uncertainty also affects whether to pursue career objectives sequentially or simultaneously. A \textit{sequential} strategy might involve maximizing $W$ first (secure finances and skills), then pivot to $M$ (impact-focused role) while maintaining sufficient $W$ and negotiating $A$. A \textit{simultaneous} strategy would seek a role that provides high $W$, $A$, and $M$ from the outset (e.g., founding a social enterprise). We hypothesize that sequential strategies dominate in expectation for risk-averse individuals with long time horizons, though this claim requires empirical validation and may not hold for individuals with exceptional comparative advantages or time-sensitive opportunities.\cite{DeLaraDoyen2008, BianchiBobba2013}

The rationale is analogous to portfolio diversification and dynamic programming: tackling one objective at a time (in a strategically ordered sequence) creates fallback options. If you first achieve a solid $W$ base, you can weather failures in later $M$ pursuits and still revert to high-$W$ roles if needed. Jumping straight into a high-$M$ startup with low $W$ may either yield a jackpot (all objectives met if the startup succeeds) or leave one with neither $W$ nor $M$ if it fails, and possibly eroded $A$ (autonomy) due to burnout or reputational loss. Unless the probability of success times its payoff strongly exceeds the safer path, a risk-adjusted utility maximizer (even one who ultimately cares about $M$) should build $W$ first.

However, simultaneous optimization may be rational for those with short time horizons or rare comparative advantages where delay significantly reduces the payoff (e.g., a timely startup idea in a fast-moving market). The framework can accommodate these calculations by assigning utility to trajectories under success/failure scenarios and comparing expected values.

\subsection{Satisficing Thresholds and Bounded Rationality}

Classical optimal career choice (maximizing a single utility function over $(W,A,M)$) is often intractable and mis-specified. Simon’s concept of \textit{satisficing} \cite{simon1956rational} offers a heuristic: define aspiration levels $W^*$, $A^*$, $M^*$ that are “good enough,” and seek any career path meeting those, rather than an unattainable global optimum.

In practice, individuals might set satisficing thresholds such as ``earn at least $X$, have control over schedule at least $Y\%$ of the time, and work on problems that alleviate issue $Z$.'' Our axes lend themselves to such thresholding: treat each $W^*$, $A^*$, $M^*$ as constraints and search for careers that satisfy all. This converts the ill-defined optimization into a constraint satisfaction problem: is there a path that meets or exceeds all three thresholds? If not, which threshold can be relaxed with least regret? This approach echoes recent work on aspirational utility and reference-dependent preferences in behavioral economics.\cite{GenicotRay2020, DaltonGhosalMani2016} 

This satisficing approach acknowledges bounded rationality: humans cannot perfectly trade off complex dimensions, but they can often judge what minimum criteria would feel acceptable. It also mitigates analysis paralysis—when faced with too many options, narrowing the search to those that clear one’s “bar” on each axis simplifies decision-making dramatically. Furthermore, thresholds can evolve: early in career one might satisfice on $M$ (accept low impact) to build $W$, then later raise $M^*$ once $W$ is sufficient, echoing the phase transitions discussed above.

In summary, embracing satisficing in multi-axis career planning encourages setting personal policies (e.g., “never go below $A$ level X after age 30” or “only consider jobs with at least moderate $M$ impact once financially independent”). Our framework provides a structure to formalize and adjust these policies over time, guiding decisions under uncertainty and cognitive constraints.

\section{Related Work and Positioning}
\label{sec:related}

We situate the Three Axes of Success framework in the context of prior research across economics, psychology, and career planning.

\paragraph{Human Capital Theory and Career Optimization:} Traditional models of career choice in labor economics, notably Becker's human capital theory \cite{becker1962investment} and Mincer's earnings equations, reduce decision-making to income maximization. Our $W$ axis builds on this foundation but extends it to include career optionality and non-wage skill value, aligning with recent critiques that pure income optimization fails to capture observed job transitions \cite{dlouhy2018path, arthur1989competing}. The notion of increasing vs.~diminishing returns in $W$ accumulation also connects to dynamic returns models (e.g., Ben-Porath's lifecycle human capital model),\cite{ben1967production} but we integrate it with non-monetary preferences.

\paragraph{Motivational Psychology:} Self-determination theory and related psychology literature emphasize autonomy and mastery as critical for motivation \cite{pink2011drive}. While these frameworks highlight the importance of $A$, they do not formalize tradeoffs with financial security ($W$) or impact ($M$). Our model quantifies Pink’s ideas (e.g., the control trap) and unites them with economic constraints. Additionally, subjective well-being studies find that beyond moderate income levels, factors like autonomy and meaning strongly predict job satisfaction \cite{kahneman2010high, rosenbaum2016money, yang2025considerations}, supporting our multi-axis approach.

\paragraph{Effective Altruism and Impact Maximization:} The 80,000 Hours paradigm \cite{todd202380k} introduced optimization of social impact in career choice, a precursor to our $M$ axis. However, its advice (e.g., priority paths, earning-to-give) has been critiqued for neglecting personal fit and risk of burnout (an $A$ and $W$ issue).\cite{Schubert2024} By embedding the EA impact metric within a broader set of objectives, we address these concerns and provide a safety valve via satisficing thresholds (ensuring, for instance, that a high-$M$ path still meets minimum $W$ for sustainability).

\paragraph{Career Construction and Narrative Identity:} Savickas' career construction theory \cite{savickas2020career} and life design counseling focus on the subjective narrative individuals craft about their careers. These qualitative approaches prioritize meaning-making and adaptability (related to $M$ and $A$) over any quantitative maximization. Our framework may appear to contrast with such models, but we view it as complementary rather than competing: the three axes can serve as a scaffold for individuals to reflect on their values and choices, structuring the narrative (``My story is about balancing financial stability with freedom and purpose'').\cite{maree2012toward} In practice, tools from life design—like envisioning alternate career selves—could map those alternatives onto $(W,A,M)$ to inform decision-making.

\paragraph{Multi-Objective Optimization and Pareto Efficiency:} Our work is also informed by decision science and operations research on multi-objective optimization.\cite{deb2016multi, kiani2019introduction} The concept of Pareto-optimal frontiers used in Section~\ref{sec:archetypes} echoes classic results in that field, but applied to personal career choices. Previous authors have modeled career decisions as multi-attribute utility problems \cite{gati1996taxonomy, savickas2020career}, though typically without the dynamic aspect we incorporate. We contribute a novel interpretation by treating career moves as traversals in a three-dimensional objective space, highlighting path dependence and non-convexities rather than static one-shot choice.

\vspace{0.5cm}

In summary, the Three Axes of Success framework synthesizes insights from diverse literatures into a cohesive model. It advances beyond income-only paradigms of economics by including autonomy and impact, grounds psychological theories of motivation in a practical decision grid, and extends effective altruism’s quantitative rigor to include personal sustainability constraints. To our knowledge, no prior work offers an explicitly three-dimensional, temporally dynamic analytic framework for career decision-making—this paper lays the groundwork for such an approach and invites further empirical and theoretical exploration.

\section{Conclusion}
\label{sec:conclusion}

We have presented a three-dimensional framework for career decision-making that formalizes Wealth, Autonomy, and Meaning as distinct yet coupled axes of success. This model enables individuals and counselors to map careers in a space that makes explicit the tradeoffs and sequential dependencies often implicit in life choices. By highlighting phenomena like control traps, adjacent possible missions, and dual-career coordination failures, the framework provides a language to diagnose why intuitive career moves can lead to frustration or underperformance when one axis is overemphasized at the expense of others.

For practitioners, the Three Axes of Success framework suggests more holistic career planning practices. Rather than asking “What job will make me happiest?” in a vague sense, individuals can ask concrete questions: “Does this option improve my $W$ sufficiently to later gain $A$ or $M$? Am I trading off too much $W$ for $M$ such that I’ll hit a financial viability limit? What minimum $A$ (control) do I need to thrive, regardless of $W$?” Such questions shift the discussion from finding a mythical optimal job to managing a portfolio of career capital, autonomy, and impact over time.

Organizations and policymakers can also derive insights. Employers seeking to retain talent might use the framework to design roles that offer a balance of financial reward, autonomy, and sense of purpose, thereby keeping employees on a locally optimal $(W,A,M)$ frontier rather than forcing a departure. Educational and professional programs could frame skill development (raising $W$) as an enabler for autonomy and impact, helping participants intentionally sequence their careers.

Several limitations warrant acknowledgment. First, the three-axis decomposition, while grounded in existing literatures, remains a modeling choice; alternative decompositions incorporating health, relationships, or social status may prove valuable in specific contexts. Second, our formalizations are stylized rather than empirically calibrated—the equations express qualitative relationships rather than quantitative predictions. Third, the framework assumes a degree of rational deliberation that may overstate how individuals actually make career decisions, particularly under stress or time pressure. Fourth, cultural and institutional variation across countries, industries, and demographic groups likely moderates many of the dynamics we describe, yet our analysis does not systematically address this heterogeneity.

With these limitations in mind, several avenues for future work emerge. Empirically, one could attempt to measure individuals' $(W,A,M)$ trajectories—using proxies like income (for $W$), job flexibility and control surveys (for $A$), and industry/social impact metrics (for $M$)—to see how real career moves cluster around the theoretical principles outlined (phase transitions, Pareto frontiers, etc.). Longitudinal datasets linking job changes to later outcomes in other dimensions would be particularly valuable.\cite{mcgonagle2012panel, vafa2022career} Qualitatively, integrating narrative methods from career construction theory could enrich the framework's applicability, ensuring that the quantitative axes do not oversimplify the complex stories and identities in vocational development.

In conclusion, careers unfold in a multidimensional space of goals and constraints. The Three Axes of Success framework offers a structured way to navigate this space, treating career design as a series of constrained optimizations and satisficing decisions under uncertainty. By making tradeoffs explicit and encouraging sequential, adaptable strategies, we aim to improve the quality of career decisions and, ultimately, the fulfillment people derive from their 80,000 hours of work.

\bibliography{03_3CDM_ref}

@article{ben1967production,
  title         = {The production of human capital and the life cycle of earnings},
  author        = {Ben-Porath, Yoram},
  journal       = {Journal of political economy},
  volume        = {75},
  number        = {4, Part 1},
  pages         = {352--365},
  year          = {1967},
  publisher     = {The University of Chicago Press},
  doi           = {10.1086/259291},
  url           = {https://doi.org/10.1086/259291}
}

@article{simon1956rational,
  title         = {Rational choice and the structure of the environment.},
  author        = {Simon, Herbert A},
  journal       = {Psychological review},
  volume        = {63},
  number        = {2},
  pages         = {129},
  year          = {1956},
  publisher     = {American Psychological Association},
  doi           = {10.1037/h0042769},
  url           = {https://doi.org/10.1037/h0042769}
}

@article{becker1962investment,
  title         = {Investment in human capital: A theoretical analysis},
  author        = {Becker, Gary S},
  journal       = {Journal of political economy},
  volume        = {70},
  number        = {5, Part 2},
  pages         = {9--49},
  year          = {1962},
  publisher     = {The University of Chicago Press},
  doi           = {10.1086/258724},
  url           = {https://doi.org/10.1086/258724}
}

@article{becker1965theory,
  title         = {A Theory of the Allocation of Time},
  author        = {Becker, Gary S},
  journal       = {The economic journal},
  volume        = {75},
  number        = {299},
  pages         = {493--517},
  year          = {1965},
  publisher     = {Oxford University Press Oxford, UK},
  doi           = {10.2307/2228949},
  url           = {https://doi.org/10.2307/2228949}
}

@article{super1973career,
  title         = {Career development theory},
  author        = {Super, Donald E and Jordaan, Jean Pierre},
  journal       = {British Journal of Guidance and Counselling},
  volume        = {1},
  number        = {1},
  pages         = {3--16},
  year          = {1973},
  publisher     = {Taylor \& Francis},
  doi           = {10.1080/03069887308259333},
  url           = {https://doi.org/10.1080/03069887308259333}
}

@article{garfield1979citation,
  title         = {Is citation analysis a legitimate evaluation tool?},
  author        = {Garfield, Eugene},
  journal       = {Scientometrics},
  volume        = {1},
  number        = {4},
  pages         = {359--375},
  year          = {1979},
  publisher     = {Akad{\'e}miai Kiad{\'o}, co-published with Springer Science+ Business Media BV},
  doi           = {10.1007/BF02019306},
  url           = {https://doi.org/10.1007/BF02019306}
}

@article{jencks1988good,
  title         = {What is a good job? A new measure of labor-market success},
  author        = {Jencks, Christopher and Perman, Lauri and Rainwater, Lee},
  journal       = {American journal of sociology},
  volume        = {93},
  number        = {6},
  pages         = {1322--1357},
  year          = {1988},
  publisher     = {University of Chicago Press},
  doi           = {10.1086/228903},
  url           = {https://doi.org/10.1086/228903}
}

@article{arthur1989competing,
  title         = {Competing technologies, increasing returns, and lock-in by historical events},
  author        = {Arthur, W Brian},
  journal       = {The economic journal},
  volume        = {99},
  number        = {394},
  pages         = {116--131},
  year          = {1989},
  publisher     = {Oxford University Press Oxford, UK},
  doi           = {10.2307/2234208},
  url           = {https://doi.org/10.2307/2234208}
}

@article{ericsson1993role,
  title         = {The role of deliberate practice in the acquisition of expert performance.},
  author        = {Ericsson, K Anders and Krampe, Ralf T and Tesch-R{\"o}mer, Clemens},
  journal       = {Psychological review},
  volume        = {100},
  number        = {3},
  pages         = {363},
  year          = {1993},
  publisher     = {American Psychological Association},
  doi           = {10.1037//0033-295X.100.3.363},
  url           = {https://graphics8.nytimes.com/images/blogs/freakonomics/pdf/DeliberatePractice(PsychologicalReview).pdf}
}

@article{kemna1993case,
  title         = {Case studies on real options},
  author        = {Kemna, Angelien GZ},
  journal       = {Financial Management},
  pages         = {259--270},
  year          = {1993},
  publisher     = {JSTOR},
  doi           = {10.2307/3665943},
  url           = {https://doi.org/10.2307/3665943}
}

@techreport{blechinger1996technological,
  title         = {Technological Change and Skill Obsolescence: The Case of German Apprenticeship Training},
  author        = {Blechinger, Doris and Pfeiffer, Friedhelm},
  year          = {1996},
  type          = {ZEW Discussion Paper},
  number        = {96-15},
  institution   = {ZEW -- Leibniz Centre for European Economic Research},
  address       = {Mannheim},
  url           = {https://www.econstor.eu/handle/10419/29394},
  note          = {URN: urn:nbn:de:bsz:31-9721}
}

@article{gati1996taxonomy,
  title         = {A taxonomy of difficulties in career decision making.},
  author        = {Gati, Itamar and Krausz, Mina and Osipow, Samuel H},
  journal       = {Journal of counseling psychology},
  volume        = {43},
  number        = {4},
  pages         = {510},
  year          = {1996},
  publisher     = {American Psychological Association},
  doi           = {10.1037/0022-0167.43.4.510},
  url           = {https://doi.org/10.1037/0022-0167.43.4.510}
}

@article{Mangematin2000,
  author        = {Mangematin, Vincent},
  title         = {{PhD} job market: professional trajectories and incentives during the {PhD}},
  journal       = {Research Policy},
  year          = {2000},
  month         = {6},
  volume        = {29},
  number        = {6},
  pages         = {741--756},
  doi           = {10.1016/S0048-7333(99)00047-5},
  url           = {https://doi.org/10.1016/S0048-7333(99)00047-5}
}

@article{jermann2002international,
  title         = {International portfolio diversification and endogenous labor supply choice},
  author        = {Jermann, Urban J},
  journal       = {European Economic Review},
  volume        = {46},
  number        = {3},
  pages         = {507--522},
  year          = {2002},
  publisher     = {Elsevier},
  doi           = {10.1016/S0014-2921(01)00149-0},
  url           = {https://doi.org/10.1016/S0014-2921(01)00149-0}
}

@article{de2002economics,
  title         = {The economics of skills obsolescence: a review},
  author        = {De Grip, Andries and Van Loo, Jasper},
  journal       = {The economics of skills obsolescence},
  pages         = {1--26},
  year          = {2002},
  publisher     = {Emerald Group Publishing Limited},
  doi           = {10.1016/S0147-9121(02)21003-1},
  url           = {https://doi.org/10.1016/S0147-9121(02)21003-1}
}

@article{ericsson2004deliberate,
  title         = {Deliberate practice and the acquisition and maintenance of expert performance in medicine and related domains},
  author        = {Ericsson, K Anders},
  journal       = {Academic medicine},
  volume        = {79},
  number        = {10},
  pages         = {S70--S81},
  year          = {2004},
  publisher     = {LWW},
  doi           = {10.1097/00001888-200410001-00022},
  url           = {https://doi.org/10.1097/00001888-200410001-00022}
}

@article{borzaga2006worker,
  title         = {Worker motivations, job satisfaction, and loyalty in public and nonprofit social services},
  author        = {Borzaga, Carlo and Tortia, Ermanno},
  journal       = {Nonprofit and voluntary sector quarterly},
  volume        = {35},
  number        = {2},
  pages         = {225--248},
  year          = {2006},
  publisher     = {Sage Publications Sage CA: Thousand Oaks, CA},
  doi           = {10.1177/0899764006287207},
  url           = {https://doi.org/10.1177/0899764006287207}
}

@article{jacobs2007real,
  title         = {Real options and human capital investment},
  author        = {Jacobs, Bas},
  journal       = {Labour Economics},
  volume        = {14},
  number        = {6},
  pages         = {913--925},
  year          = {2007},
  publisher     = {Elsevier},
  doi           = {10.1016/j.labeco.2007.06.008},
  url           = {https://doi.org/10.1016/j.labeco.2007.06.008}
}

@inbook{DeLaraDoyen2008,
  author        = {De Lara, Michel and Doyen, Luc},
  title         = {Sequential decision under imperfect information},
  booktitle     = {Sustainable Management of Natural Resources},
  publisher     = {Springer},
  year          = {2008},
  series        = {Environmental Science and Engineering},
  chapter       = {9},
  pages         = {221--236},
  doi           = {10.1007/978-3-540-79074-7_9},
  url           = {https://doi.org/10.1007/978-3-540-79074-7_9}
}

@article{kerr2010supply,
  title         = {The supply side of innovation: H-1B visa reforms and US ethnic invention},
  author        = {Kerr, William R and Lincoln, William F},
  journal       = {Journal of Labor Economics},
  volume        = {28},
  number        = {3},
  pages         = {473--508},
  year          = {2010},
  publisher     = {The University of Chicago Press},
  doi           = {10.1086/651934},
  url           = {https://www.hbs.edu/ris/Publication%20Files/09-005_005359f2-2ee8-4d73-b248-af492e44ecb4.pdf}
}

@article{kahneman2010high,
  title         = {High income improves evaluation of life but not emotional well-being},
  author        = {Kahneman, Daniel and Deaton, Angus},
  journal       = {Proceedings of the national academy of sciences},
  volume        = {107},
  number        = {38},
  pages         = {16489--16493},
  year          = {2010},
  publisher     = {National Academy of Sciences},
  doi           = {10.1073/pnas.1011492107},
  url           = {https://doi.org/10.1073/pnas.1011492107}
}

@book{pink2011drive,
  title         = {Drive: The Surprising Truth About What Motivates Us},
  author        = {Pink, Daniel H.},
  year          = {2011},
  publisher     = {Riverhead Books},
  address       = {New York, NY},
  isbn          = {978-1594484803},
  note          = {Originally published in 2009}
}

@article{berglann2011entrepreneurship,
  title         = {Entrepreneurship: Origins and returns},
  author        = {Berglann, Helge and Moen, Espen R and R{\o}ed, Knut and Skogstr{\o}m, Jens Fredrik},
  journal       = {Labour economics},
  volume        = {18},
  number        = {2},
  pages         = {180--193},
  year          = {2011},
  publisher     = {Elsevier},
  doi           = {10.1016/j.labeco.2010.10.002},
  url           = {https://doi.org/10.1016/j.labeco.2010.10.002}
}

@article{peri2012effect,
  title         = {The effect of immigration on productivity: Evidence from US states},
  author        = {Peri, Giovanni},
  journal       = {Review of Economics and Statistics},
  volume        = {94},
  number        = {1},
  pages         = {348--358},
  year          = {2012},
  publisher     = {The MIT Press},
  doi           = {10.1162/REST_a_00137},
  url           = {https://doi.org/10.1162/REST_a_00137}
}

@misc{mmm2012math,
  author        = {Adeney, Peter},
  title         = {The Shockingly Simple Math Behind Early Retirement},
  howpublished  = {Mr. Money Mustache Blog},
  year          = {2012},
  month         = jan,
  day           = {13},
  url           = {https://www.mrmoneymustache.com/2012/01/13/the-shockingly-simple-math-behind-early-retirement/},
  note          = {Accessed: January 14, 2026. Popularized the 4\% rule and high savings rate framework.}
}

@article{maree2012toward,
  title         = {Toward a Combined Qualitative-Quantitative Approach: Advancing Postmodern Career Counselling Theory and Practice},
  author        = {Maree, Jacobus G. and Morgan, Brandon},
  journal       = {Cypriot Journal of Educational Sciences},
  volume        = {7},
  number        = {4},
  pages         = {311--325},
  year          = {2012},
  issn          = {1305-905X},
  url           = {https://un-pub.eu/ojs/index.php/cjes/article/view/915},
  note          = {Published by Academic World Education and Research Center}
}

@article{mcgonagle2012panel,
  title         = {The Panel Study of Income Dynamics: Overview, Recent Innovations, and Potential for Life Course Research},
  author        = {McGonagle, Katherine A. and Schoeni, Robert F. and Sastry, Narayan and Freedman, Vicki A.},
  journal       = {Longitudinal and Life Course Studies},
  volume        = {3},
  number        = {2},
  pages         = {268--284},
  year          = {2012},
  doi           = {10.14301/llcs.v3i2.188},
  pmid          = {23482334},
  pmcid         = {PMC3591471},
  url           = {https://psidonline.isr.umich.edu/llcs2012.pdf},
  note          = {NIHMSID: NIHMS393237}
}

@article{BianchiBobba2013,
  author  = {Bianchi, Milo and Bobba, Matteo},
  title   = {Liquidity, Risk, and Occupational Choices},
  journal = {The Review of Economic Studies},
  year    = {2013},
  volume  = {80},
  number  = {2},
  pages   = {491--511},
  month   = {4},
  doi     = {10.1093/restud/rds031},
  url           = {https://doi.org/10.1093/restud/rds031}
}

@incollection{hartung2013life,
  author        = {Hartung, Paul J.},
  title         = {The Life-Span, Life-Space Theory of Careers},
  booktitle     = {Career Development and Counseling: Putting Theory and Research to Work},
  editor        = {Brown, Steven D. and Lent, Robert W.},
  edition       = {2nd},
  publisher     = {John Wiley \& Sons},
  address       = {Hoboken, NJ},
  year          = {2013},
  pages         = {83--113},
  isbn          = {978-1118063354}
}

@book{newport2016so,
  title         = {So Good They Can't Ignore You: Why Skills Trump Passion in the Quest for Work You Love},
  author        = {Newport, Cal},
  year          = {2016},
  publisher     = {Business Plus},
  address       = {London, UK},
  isbn          = {978-0349415864},
  note          = {Originally published in 2012}
}

@article{rosenbaum2016money,
  title         = {Money isn't everything: Job satisfaction, nonmonetary job rewards, and sub-baccalaureate credentials},
  author        = {Rosenbaum, Janet E. and Rosenbaum, James E.},
  journal       = {Research in Higher Education Journal},
  volume        = {30},
  pages         = {1--21},
  month         = {sep},
  year          = {2016},
  issn          = {1941-3432},
  pmid          = {31080844},
  pmcid         = {PMC6508652},
  url           = {https://www.aabri.com/manuscripts/162430.pdf},
  note          = {Manuscript ID: 162430},
  publisher     = {Academic and Business Research Institute}
}

@article{mcmahan2016philosophical,
  title         = {Philosophical critiques of effective altruism},
  author        = {McMahan, Jeff},
  journal       = {Philosophers' Magazine},
  volume        = {73},
  year          = {2016},
  publisher     = {Philosophy Documentation Center},
  doi           = {10.5840/tpm20167379},
  url           = {https://dx.doi.org/10.5840/tpm20167379}
}

@article{blundell2016consumption,
  title         = {Consumption inequality and family labor supply},
  author        = {Blundell, Richard and Pistaferri, Luigi and Saporta-Eksten, Itay},
  journal       = {American Economic Review},
  volume        = {106},
  number        = {2},
  pages         = {387--435},
  year          = {2016},
  publisher     = {American Economic Association 2014 Broadway, Suite 305, Nashville, TN 37203},
  doi           = {10.1257/aer.20121549},
  url           = {https://dx.doi.org/10.1257/aer.20121549}
}

@incollection{deb2016multi,
  author        = {Deb, Kalyanmoy and Sindhya, Karthik and Hakanen, Jussi},
  title         = {Multi-objective Optimization},
  booktitle     = {Decision Sciences: Theory and Practice},
  editor        = {Raghavendra, Raghu Nandan Sengupta and Gupta, Aparna Kumar and Dutta, Joydeep},
  year          = {2016},
  pages         = {161--200},
  publisher     = {CRC Press},
  address       = {Boca Raton, FL},
  doi           = {10.1201/9781315183176-12},
  isbn          = {978-1315183176},
  url           = {https://www.taylorfrancis.com/chapters/edit/10.1201/9781315183176-12/multi-objective-optimization-kalyanmoy-deb-karthik-sindhya-jussi-hakanen}
}

@article{DaltonGhosalMani2016,
  author  = {Dalton, Patricio S. and Ghosal, Sayantan and Mani, Anandi},
  title   = {Poverty and Aspirations Failure},
  journal = {The Economic Journal},
  year    = {2016},
  volume  = {126},
  number  = {590},
  pages   = {165--188},
  month   = {2},
  doi     = {10.1111/ecoj.12210},
  url     = {https://doi.org/10.1111/ecoj.12210}
}

@inbook{nagele2016education,
  author    = {N{\"a}gele, Christof and Stalder, Barbara E.},
  title     = {Competence and the Need for Transferable Skills},
  booktitle = {Competence-based Vocational and Professional Education},
  publisher = {Springer},
  year      = {2016},
  series    = {Technical and Vocational Education and Training: Issues, Concerns and Prospects},
  volume    = {23},
  chapter   = {34},
  pages     = {739--753},
  doi       = {10.1007/978-3-319-41713-4_34},
  url       = {https://link.springer.com/chapter/10.1007/978-3-319-41713-4_34}
}

@article{lopes2017determinants,
  title         = {The determinants of work autonomy and employee involvement: A multilevel analysis},
  author        = {Lopes, Helena and Calapez, Teresa and Lopes, Diniz},
  journal       = {Economic and Industrial Democracy},
  volume        = {38},
  number        = {3},
  pages         = {448--472},
  year          = {2017},
  publisher     = {SAGE Publications Sage UK: London, England},
  doi           = {10.1177/0143831X15579226},
  url           = {https://doi.org/10.1177/0143831X15579226}
}

@article{dlouhy2018path,
  title         = {Path dependence in occupational careers: Understanding occupational mobility development throughout individuals' careers},
  author        = {Dlouhy, Katja and Biemann, Torsten},
  journal       = {Journal of Vocational Behavior},
  volume        = {104},
  pages         = {86--97},
  year          = {2018},
  publisher     = {Elsevier},
  doi           = {10.1016/j.jvb.2017.10.009},
  url           = {https://doi.org/10.1016/j.jvb.2017.10.009}
}

@article{cantamessa2018startups,
  title         = {Startups' roads to failure},
  author        = {Cantamessa, Marco and Gatteschi, Valentina and Perboli, Guido and Rosano, Mariangela},
  journal       = {Sustainability},
  volume        = {10},
  number        = {7},
  pages         = {2346},
  year          = {2018},
  publisher     = {MDPI},
  doi           = {10.3390/su10072346},
  url           = {https://doi.org/10.3390/su10072346}
}

@incollection{kiani2019introduction,
  author        = {Kiani-Moghaddam, Mohammad and Shivaie, Mojtaba and Weinsier, Philip D.},
  title         = {Introduction to Multi-objective Optimization and Decision-Making Analysis},
  booktitle     = {Modern Music-Inspired Optimization Algorithms for Electric Power Systems},
  editor        = {Shivaie, Mojtaba and Weinsier, Philip D.},
  series        = {Springer Optimization and Its Applications},
  volume        = {142},
  pages         = {21--45},
  year          = {2019},
  publisher     = {Springer Nature},
  address       = {Cham, Switzerland},
  doi           = {10.1007/978-3-030-12044-3_2},
  isbn          = {978-3-030-12043-6},
  url           = {https://link.springer.com/chapter/10.1007/978-3-030-12044-3_2}
}

@incollection{savickas2020career,
  author        = {Savickas, Mark L.},
  title         = {Career Construction Theory and Counseling Model},
  booktitle     = {Career Development and Counseling: Putting Theory and Research to Work},
  editor        = {Brown, Steven D. and Lent, Robert W.},
  edition       = {3rd},
  year          = {2020},
  pages         = {165--200},
  publisher     = {John Wiley \& Sons},
  address       = {Hoboken, NJ},
  doi           = {10.1002/9781394258994.ch6},
  isbn          = {978-1119582205},
  url           = {https://onlinelibrary.wiley.com/doi/abs/10.1002/9781394258994.ch6}
}

@article{ellul2020career,
  title         = {Career risk and market discipline in asset management},
  author        = {Ellul, Andrew and Pagano, Marco and Scognamiglio, Annalisa},
  journal       = {The Review of Financial Studies},
  volume        = {33},
  number        = {2},
  pages         = {783--828},
  year          = {2020},
  publisher     = {Oxford University Press},
  doi           = {10.1093/rfs/hhz062},
  url           = {https://doi.org/10.1093/rfs/hhz062}
}

@misc{socap2020dualcareer,
  author        = {{SOCAP Global}},
  title         = {Dual-Career Couples Are the New Norm: What Business Leaders Need to Know},
  year          = {2020},
  month         = jan,
  day           = {16},
  url           = {https://socapglobal.com/2020/01/dual-career-couples-are-the-new-norm-what-business-leaders-need-to-know/},
  howpublished  = {SOCAP Global News},
  note          = {In collaboration with Berkeley Haas Center for Equity, Gender, and Leadership (EGAL); Accessed: January 14, 2026}
}

@article{GenicotRay2020,
  author  = {Genicot, Garance and Ray, Debraj},
  title   = {Aspirations and Economic Behavior},
  journal = {Annual Review of Economics},
  year    = {2020},
  volume  = {12},
  pages   = {715--746},
  month   = {8},
  doi     = {10.1146/annurev-economics-080217-053245},
  url     = {https://doi.org/10.1146/annurev-economics-080217-053245}
}

@techreport{peetz2021living,
  author        = {Peetz, Johanna and Robson, Jennifer},
  title         = {Living Gig to Gig and Paycheque to Paycheque: How Income Volatility Affects Financial Decisions},
  institution   = {Center for Economic Policy Research (CEPR)},
  year          = {2021},
  type          = {Report},
  url           = {https://cepr.org/system/files/2022-08/Living%20Gig%20to%20Gig%20and%20Paycheque%20to%20Paycheque%20How%20Income%20Volatility%20Affects%20Financial%20Decisions%20-%20Johanna%20Peetz%20and%20Jennifer%20Robson.pdf},
  note          = {Research report funded by the Government of Canada}
}

@article{mei2021does,
  title         = {Does practice enhance adaptability? The role of personality trait, supervisor behavior, and career development training},
  author        = {Mei, Mei and Yang, Fu and Tang, Mingfeng},
  journal       = {Frontiers in Psychology},
  volume        = {11},
  pages         = {594791},
  year          = {2021},
  publisher     = {Frontiers Media SA},
  doi           = {10.3389/fpsyg.2020.594791},
  url           = {https://doi.org/10.3389/fpsyg.2020.594791}
}

@article{chen2021relationship,
  title         = {The Relationship Between Intolerance of Uncertainty and Employment Anxiety of Graduates During {COVID}-19: The Moderating Role of Career Planning},
  author        = {Chen, Li and Zeng, Shuyu},
  journal       = {Frontiers in Psychology},
  volume        = {12},
  pages         = {694785},
  year          = {2021},
  month         = {oct},
  doi           = {10.3389/fpsyg.2021.694785},
  url           = {https://doi.org/10.3389/fpsyg.2021.694785},
  pmid          = {34764900},
  pmcid         = {PMC8576396},
  publisher     = {Frontiers Media SA},
  issn          = {1664-1078}
}

@techreport{rueda2021career,
  title         = {Career Paths with a Two-Body Problem: Occupational Specialization and Geographic Mobility},
  author        = {Rueda, Valeria and Wilemme, Guillaume},
  year          = {2021},
  institution   = {W.E. Upjohn Institute for Employment Research},
  type          = {Working Paper},
  number        = {21-346},
  doi           = {10.17848/wp21-346},
  url           = {https://doi.org/10.17848/wp21-346},
  address       = {Kalamazoo, MI}
}

@article{berkey2021philosophical,
  title         = {The philosophical core of effective altruism},
  author        = {Berkey, Brian},
  journal       = {Journal of Social Philosophy},
  volume        = {52},
  number        = {1},
  pages         = {93--115},
  year          = {2021},
  doi           = {10.1111/josp.12347},
  url           = {https://doi.org/10.1111/josp.12347}
}

@incollection{yates2022getting,
  author        = {Yates, Tyler P.},
  title         = {Getting ``Us'' a Job: The Two+ Body Problem and the Academic Job Market},
  booktitle     = {Strategies for Navigating Graduate School and Beyond},
  editor        = {Lorentz, II, Kevin G. and Mallinson, Daniel J. and Hellwege, Julia Marin and Phoenix, Davin L. and Strachan, J. Cherie},
  year          = {2022},
  chapter       = {48},
  pages         = {324--330},
  publisher     = {American Political Science Association},
  address       = {Washington, DC},
  isbn          = {978-1878147745},
  url           = {https://apsanet.org/wp-content/uploads/2024/08/GS1-Chapter-48.pdf}
}

@article{zhang2022teacher,
  title         = {Teacher burnout and turnover intention in higher education: The mediating role of job satisfaction and the moderating role of proactive personality},
  author        = {Zhang, Qun and Li, Xianyin and Gamble, Jeffrey Hugh},
  journal       = {Frontiers in Psychology},
  volume        = {13},
  pages         = {1076277},
  year          = {2022},
  publisher     = {Frontiers Media SA},
  doi           = {10.3389/fpsyg.2022.1076277},
  url           = {https://doi.org/10.3389/fpsyg.2022.1076277}
}

@article{nie2023job,
  title         = {Job autonomy and work meaning: Drivers of employee job-crafting behaviors in the VUCA times},
  author        = {Nie, Ting and Tian, Min and Cai, Mingyang and Yan, Qiao},
  journal       = {Behavioral Sciences},
  volume        = {13},
  number        = {6},
  pages         = {493},
  year          = {2023},
  publisher     = {MDPI},
  doi           = {10.3390/bs13060493},
  url           = {https://doi.org/10.3390/bs13060493}
}

@article{tam2023migration,
  title         = {Migration of dual-earner couples: a subjective wellbeing approach},
  author        = {Tam, Diana and Grimes, Arthur},
  journal       = {Review of Economics of the Household},
  volume        = {21},
  number        = {1},
  pages         = {269--289},
  year          = {2023},
  publisher     = {Springer},
  doi           = {10.1007/s11150-021-09598-z},
  url           = {https://doi.org/10.1007/s11150-021-09598-z}
}

@article{zhang2023impacts,
  title         = {The impacts of altruism levels on the job preferences of medical students: a cross-sectional study in China},
  author        = {Zhang, Yue and Lin, Xing and Li, Xing and Han, Youli},
  journal       = {BMC medical education},
  volume        = {23},
  number        = {1},
  pages         = {538},
  year          = {2023},
  publisher     = {Springer},
  doi           = {10.1186/s12909-023-04490-z},
  url           = {https://doi.org/10.1186/s12909-023-04490-z}
}

@misc{express2023salary,
  author        = {{Express Employment Professionals}},
  title         = {Job Seekers Willing to Sacrifice Salary for Freedom},
  howpublished  = {Express Employment Professionals Newsroom},
  year          = {2023},
  month         = oct,
  day           = {11},
  url           = {https://www.expresspros.com/newsroom/news-releases/news-releases/2023/10/job-seekers-willing-to-sacrifice-salary-for-freedom},
  note          = {Accessed: January 14, 2026}
}

@misc{bls2023software,
  author        = {{U.S. Bureau of Labor Statistics}},
  title         = {Software Developers, Quality Assurance Analysts, and Testers: Occupational Outlook Handbook},
  year          = {2023},
  howpublished  = {U.S. Department of Labor},
  url           = {https://www.bls.gov/ooh/computer-and-information-technology/software-developers.htm},
  note          = {Accessed: January 14, 2026}
}

@misc{lightcast2023labor,
  author        = {{Lightcast}},
  title         = {Labor Insight\texttrademark Real-Time Labor Market Information},
  year          = {2023},
  howpublished  = {Lightcast Database},
  url           = {https://lightcast.io/},
  note          = {Formerly Burning Glass Technologies; Accessed: January 14, 2026}
}

@book{todd202380k,
  title         = {80,000 Hours: Find a Fulfilling Career That Does Good},
  author        = {Todd, Benjamin},
  year          = {2023},
  edition       = {4th},
  publisher     = {80,000 Hours},
  address       = {Oxford, UK},
  isbn          = {978-1537324005},
  url           = {https://80000hours.org/book/},
  note          = {Originally published in 2016; updated regularly as an online guide and print edition.}
}

@article{avlonitis2023career,
  title         = {Career path recommendations for long-term income maximization: A reinforcement learning approach},
  author        = {Avlonitis, Spyros and Lavi, Dor and Mansoury, Masoud and Graus, David},
  journal       = {arXiv preprint arXiv:2309.05391},
  year          = {2023},
  doi           = {10.48550/arXiv.2309.05391},
  url           = {https://doi.org/10.48550/arXiv.2309.05391}
}

@techreport{burningglassinstitute2023compass,
  author        = {{The Burning Glass Institute}},
  title         = {2023 Skills Compass Report: Strategic Skills Planning for Upskilling and Reskilling the Workforce},
  institution   = {The Burning Glass Institute},
  year          = {2023},
  url           = {https://www.burningglassinstitute.org/research/2023-skills-compass-report},
  note          = {In cooperation with Lightcast data regarding industry transferability matrices}
}

@incollection{Schubert2024,
  author        = {Stefan Schubert and Lucius Caviola},
  title         = {Effective Altruism for Mortals},
  booktitle     = {Effective Altruism and the Human Mind: The Clash Between Impact and Intuition},
  publisher     = {Oxford University Press},
  year          = {2024},
  chapter       = {9},
  doi           = {10.1093/oso/9780197757376.003.0010},
  url           = {https://doi.org/10.1093/oso/9780197757376.003.0010}
}

@article{averkamp2024decomposing,
  title         = {Decomposing gender wage gaps: a family economics perspective},
  author        = {Averkamp, Doroth{\'e}e and Bredemeier, Christian and Juessen, Falko},
  journal       = {The Scandinavian Journal of Economics},
  volume        = {126},
  number        = {1},
  pages         = {3--37},
  year          = {2024},
  publisher     = {Wiley Online Library},
  doi           = {10.1111/sjoe.12542},
  url           = {https://doi.org/10.1111/sjoe.12542}
}

@article{vafa2022career,
  title         = {CAREER: A foundation model for labor sequence data},
  author        = {Vafa, Keyon and Palikot, Emil and Du, Tianyu and Kanodia, Ayush and Athey, Susan and Blei, David M},
  journal       = {arXiv preprint arXiv:2202.08370},
  year          = {2022},
  doi           = {10.48550/arXiv.2202.08370},
  url           = {https://doi.org/10.48550/arXiv.2202.08370}
}

@misc{levelsfyi2024frontier,
  author        = {{Levels.fyi}},
  title         = {Frontier Technologies Salaries},
  year          = {2024},
  url           = {https://www.levels.fyi/companies/frontier-technologies/salaries},
  note          = {Total compensation data via skill ladders; Accessed: January 14, 2026}
}

@techreport{vanguard2024has,
  author        = {{Vanguard Group}},
  title         = {How America Saves 2024},
  institution   = {Vanguard Strategy \& Research},
  year          = {2024},
  month         = jun,
  url           = {https://workplace.vanguard.com/content/dam/inst/iig-transformation/insights/pdf/2024/How_America_Saves_2024_Early_Preview.pdf},
  note          = {23rd Edition; Analysis of 2023 participant behavior and liquid savings distributions}
}

@article{mello2025career,
  title         = {Career Capital Development in Global Work: The Roles of Job Scope, Career Adaptability, and Gender Dynamics},
  author        = {Mello, Rodrigo and Suutari, Vesa and Kemppinen, Samu},
  journal       = {European Management Journal},
  volume        = {43},
  number        = {2},
  year          = {2025},
  month         = apr,
  issn          = {0263-2373},
  doi           = {10.1016/j.emj.2025.04.001},
  url           = {https://doi.org/10.1016/j.emj.2025.04.001},
  publisher     = {Elsevier},
  address       = {Vaasa, Finland}
}

@article{yang2025considerations,
  title         = {Considerations beyond salary: study of job satisfaction among Chinese social work practitioners in different positions},
  author        = {Yang, Hui and Lv, Zhezhen and Zhang, Ying and Cui, Ting and Chen, Honglin},
  journal       = {Asia Pacific Journal of Social Work and Development},
  volume        = {35},
  number        = {2},
  pages         = {192--211},
  year          = {2025},
  publisher     = {Taylor \& Francis},
  doi           = {10.1080/29949769.2024.2374295},
  url           = {https://doi.org/10.1080/29949769.2024.2374295}
}

@article{fernandez2025gender,
  title         = {Gender diversity and innovation performance in family firms: Evidence from the Spanish manufacturing industry},
  author        = {Fern{\'a}ndez-L{\'o}pez, Sara and Rodr{\'\i}guez-Gul{\'\i}as, Mar{\'\i}a Jes{\'u}s and Calvo, Nuria and Rodeiro-Pazos, David},
  journal       = {Journal of Engineering and Technology Management},
  volume        = {77},
  pages         = {101904},
  year          = {2025},
  publisher     = {Elsevier},
  doi           = {10.1016/j.jengtecman.2025.101904},
  url           = {https://doi.org/10.1016/j.jengtecman.2025.101904}
}

@article{yu2025dollars,
  title         = {Dollars and Domestic Duties: A 22-Year Study of Income, Home Labor, and Gendered Career Outcomes in Dual-Earner Couples},
  author        = {Yu, Hyejin and Smith, Alexis Nicole and Dimotakis, Nikolaos},
  journal       = {Journal of Organizational Behavior},
  volume        = {46},
  number        = {5},
  pages         = {662--684},
  year          = {2025},
  publisher     = {Wiley Online Library},
  doi           = {10.1002/job.2879},
  url           = {https://doi.org/10.1002/job.2879}
}

@misc{goodlifejourney2025multi,
  author        = {{The Good Life Journey}},
  title         = {Sequential Careers \& FI: How to Design Multiple Career Paths},
  year          = {2025},
  month         = {Nov 26},
  url           = {https://www.thegoodlifejourney.com/home/multi-career-life-sequential-careers-financial-independence},
  note          = {Accessed: 2026-01-14}
}

@misc{bipventures2025powerlaw,
  author        = {{BIP Ventures}},
  title         = {Explainer: What is the Venture Capital Power Law?},
  howpublished  = {BIP Ventures News},
  year          = {2025},
  month         = apr,
  day           = {16},
  url           = {https://www.bipventures.vc/news/explainer-what-is-the-venture-capital-power-law},
  note          = {Accessed: January 15, 2026}
}

@article{casalhay2025gig,
  author        = {Casalhay, Sharmaine F. and Guevarra, Carla Marie and Bragas, Cresilda M.},
  title         = {The Gig Economy: Financial Challenges and Opportunities Faced by Freelancers},
  journal       = {International Journal of Research Publication and Reviews},
  year          = {2025},
  volume        = {6},
  number        = {5},
  pages         = {3545--3560},
  month         = {May},
  doi           = {10.55248/gengpi.6.0525.1716},
  url           = {https://doi.org/10.55248/gengpi.6.0525.1716}
}

@techreport{BenjaminMilouchevaVigezzi2025,
  author        = {Benjamin, Dwayne and Miloucheva, Boriana and Vigezzi, Natalia},
  title         = {The Opportunity Cost of a {PhD}: Spending your Twenties},
  institution   = {University of Toronto, Department of Economics},
  year          = {2025},
  type          = {Working Paper},
  number        = {802},
  url           = {https://www.economics.utoronto.ca/public/workingPapers/tecipa-802.pdf}
}

\end{document}